\documentclass[a4paper,prb]{revtex4}
\usepackage{graphicx}          
\usepackage{dcolumn}
\usepackage{amssymb,amsmath,color}

\begin{document}
\title[Short Title]{Microfabrication of Three-Dimensional Structures in Polymer and Glass by Femtosecond
Pulses}
\thanks{This is a proceedings paper of bi-lateral Conf. (Republics of China \& Lithuania) on Optoelectronics and Magnetic Materials, Taipei, May 25-26, 2002.}

\author{Saulius Juodkazis}
\thanks{Corresponding author.\\\protect S.J., V.M, and E.V. also, at Institute of Materials Science and Applied Research,
Vilnius University, Saul\.{e}tekio 10, 2040 Vilnius, Lithuania}
\email{saulius@eco.tokushima-u.ac.jp. (fax: (+81) 88 656 7598)}
\author{Toshiaki Kondo}
\author{Vygantas Mizeikis}
\author{Shigeki Matsuo}
\author{Hiroaki Misawa}
\affiliation{Core Research for Evolution Science \& Technology
(CREST), Japan Science \& Technology Corporation (JST), The
University of Tokushima, 2-1 Minamijyosanjima, Tokushima 770-8506,
Japan}
\author{Egidijus Vanagas}
\author{Igor Kudryashov}
\affiliation{Tokyo Instruments Inc., 6-18-14 Nishikasai, Edogawa-ku,
Tokyo 134-0088, Japan}

\date{\today}


\begin{abstract}
We report three-dimensional laser microfabrication, which enables
microstructuring of materials on the scale of $0.2-1~\mu$m. The two
different types of microfabrication demonstrated and discussed in
this work are based on holographic recording, and light-induced
damage in transparent dielectric materials. Both techniques use
nonlinear optical excitation of materials by ultrashort laser pulses
(duration $< 1$~ps).
\end{abstract}

\pacs{direct laser writing, femtosecond microfabrication, photonic
crystals, 3D optical memory}

\maketitle

\section[Introduction]{Introduction}

Technology used for the production of semiconductor microchips will
soon enable industry-grade fabrication with smallest feature size of
100~nm. The new blue-laser DVD format, agreed upon in February 2002,
will feature up to 27~GBytes of memory on one side of a single 12-cm
disc, nearly six times the capacity of current 4.7~GBytes disks.
However, these are the achievements of 2D microfabrication.
Developing technologies of 3D fabrication which would enable to
achieve minimum feature size of $0.1-1~\mu$m  is still a challenge,
and attracts increasing interest. Tools for manipulation, handling,
and fabrication on this scale are important for the future of
microfabrication and microassembling techniques. The interest is also
prompted by the biocompatibility issues, since the most crucial
processes in living organisms occur on the length scale of
$0.1-1~\mu$m. The early works on 3D patterning were focused on the
formation of ordered ``optical matter'' structures composed of
microspheres~\cite{Burns749} suspended in liquid solutions. The
ordering was accomplished via the force exerted on the microspheres
by periodical multi-beam light interference field~\cite{Burns749}.
Recently, we reported on fabrication of 3D structures in
resist~\cite{Kondo725} by using diffractive beam splitter and
objective lens focusing, which were used to create a periodic
pattern. 3D fabrication was achieved via one-photon absorption at
400~nm in the resist.

\begin{figure}[b]
\begin{center}
\includegraphics[angle=-90,width=16.20cm]{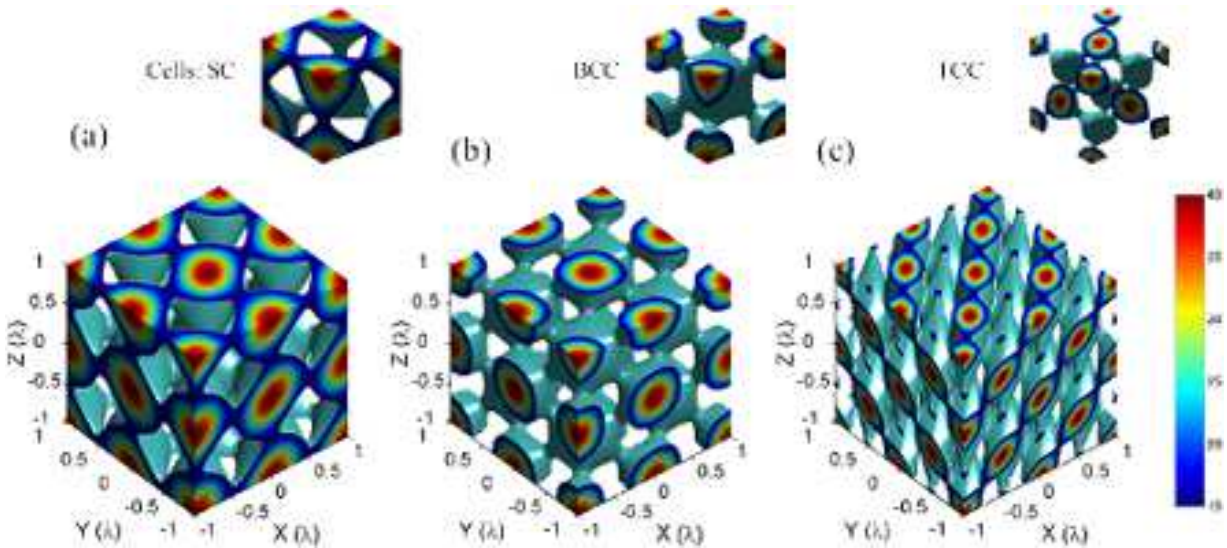}
\caption{\small{Light intensity distribution corresponding to simple
cubic (SC)(a), body-centered cubic (BCC)(b), and  face centered cubic
(FCC)(c) lattices generated by four interfering beams}. The
coordinates are given in the units normalized to the wavelength
$\lambda$.}\label{f-cai}
\end{center}
\end{figure}

In this work we report a fabrication via two-photon absorption of
800~nm irradiation in resist. Such approach may be particularly
useful for the fabrication of photonic crystals (PhC). It can be
shown that any 3D Bravais lattice can be generated by the
interference of four non-coplanar beams~\cite{Cai}. This is
illustrated in Fig.~\ref{f-cai} which shows calculated light
intensity distributions having cubic lattice symmetry. In addition to
such \emph{holographic recording}, we also report 3D microstructuring
by \emph{direct laser writing}~\cite{Miwa}. This method is based on
the light-induced damaging of materials. High irradiance laser
pulses, tightly focused by high numerical aperture (NA) optics,
induce permanent damage inside the bulk of optically transparent
dielectrics; the size of the damaged region is usually smaller than
1~$\mu$m owing to the tight focusing and nonlinear nature of the
material excitation. The prove of a principle for Tbits/cm$^{3}$
memory fabrication is demonstrated.

\section[Experimental]{Experimental}

\begin{figure}[t]
\begin{center}
\includegraphics[angle=-90,width=10.0cm,keepaspectratio=true]{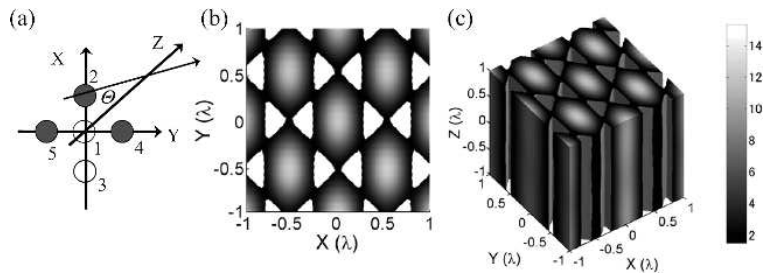}
\caption{\small{(a) Laser beam arrangement in holographic fabrication
experiments. The beams used in experiments and calculations with
three beam interference are indicated by the filled circles. (b)
calculated pattern of the light intensity distribution in the
XY-plane for the intensities which exceed the threshold intensity for
photoresist modification by the factor of 1.5. (c) calculated 3D
light intensity distribution in the beam intersection. Dimensions of
all images are $2\lambda \times 2\lambda \times2\lambda$ along the
$XYZ$ directions. All incident beams have equal intensity of 1 and
phase 0. The angle $\theta=69^{\circ}$ was taken for calculations.}}
\label{f-3b}
\end{center}
\end{figure}
\begin{figure}[t]
\begin{center}
\includegraphics[angle=-90,width=15.50cm,keepaspectratio=true]{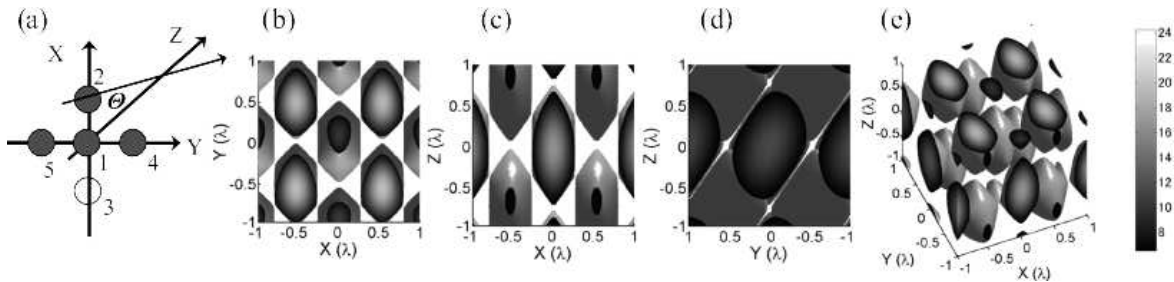}
\caption{\small{(a) The geometry of holographic fabrication with four
beams. (b-d) Light intensity distribution in the XY-, XZ, and
YZ-planes, respectively. The conditions are the same as in
Fig.~\ref{f-3b}. (e) 3D light intensity distribution at the beam
intersection. The constant intensity surface was chosen to be at 6.5,
which corresponds to unconnected structure.}} \label{f-4b}
\end{center}
\end{figure}

Holographic recording experiments were carried out using 150~fs
duration pulses derived from a Ti:sapphire laser at the fundamental
800~nm wavelength. Multiple coherent beams were obtained from a
single beam by using a diffractive beam splitter. For the direct
laser writing, amplified pulses of fundamental wavelength from the
same laser were used. Beam focusing was achieved by $NA>1$ microscope
objective lens. More details concerning the setup can be found
elsewhere~\cite{Kondo725,Miwa,Watanabe}. We have used commercial SU-8
photoresist (sensitive at 400~nm wavelength) for the holographic
recording, and fused silica (transparent at 800~nm wavelength) for
the direct laser writing experiments.

\section[Results and Discussion]{Results and Discussion}
\subsection[Holographic recording]{Holographic recording}

Prior to describing the experimental results, it is helpful to
demonstrate theoretically the possibilities to obtain different 3D as
well as 2D light interference patterns by using various numbers of
plane waves with certain amplitude and phase. The simplest 2D
structures can be fabricated by the three side beams (the central
beams blocked) as shown in Fig.~\ref{f-3b}(a). The developed
structure should be self-supporting, i.e., consist of well connected
regions. An example of such structure is shown in Fig.~\ref{f-3b}(b),
where the rod-like high intensity regions are joined into the
self-supporting structure at the light intensity threshold of 1.5.
For the PhC applications it may be important to control the volume
fraction of the unexposed photoresist, later to be removed in the
development process.  This can be achieved by adjusting the exposure.
When the central beam is turned on, 3D interference patterns may
result  as illustrated in Fig.~\ref{f-4b}).

\begin{figure}[t]
\begin{center}
\includegraphics[angle=-90,width=16.50cm,keepaspectratio=true]{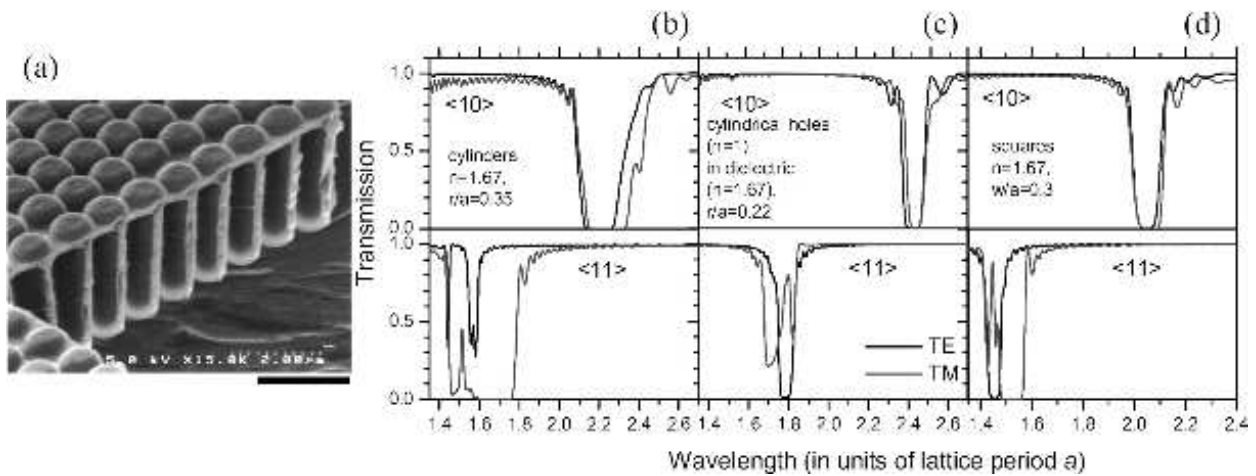}
\caption{\small{(a) SEM image of the structure fabricated by four
beam holographic technique (beams \emph{2,3,4}, and \emph{5} shown in
Fig.~\ref{f-3b}(a) were used). The exposure time at 1~kHz repetition
rate was 120~s. The focusing of beams corresponded to an
$\theta=42^{\circ}$ angle with Z-axis. Scale bar is $2~\mu$m. (b-d)
Simulated transmission spectra of the 2D photonic crystal shown
in~(a). TE and TM denote two orthogonal linear polarizations of the
electromagnetic waves, with TE being parallel to the rods/cylinders.
The propagation direction labeled $\langle 10\rangle$ is along the
sides of the primitive square cell, while $\langle 11\rangle$ is
along its diagonals.}} \label{f-trans}
\end{center}
\end{figure}

Image of the 2D pillar structure fabricated by the holographic
technique is shown in Fig.~\ref{f-trans}(a). The recording wavelength
used was 800~nm, at which two-photon absorption was required for the
photomodification of the photoresist. The thickness of the
photoresist film was $4~\mu$m. High sample quality and periodicity is
evident from the figure. In such structure spatial modulation of the
dielectric constant may result in the formation of photonic bands,
and therefore it is potentially applicable as a PhC. There is a
growing interest in developing novel techniques of the PhC
fabrication~\cite{Mizeikis35}, and the holographic approach described
above is potentially interesting in this respect. Simulated
transmission spectra of the 2D PhC are shown in
Fig.~\ref{f-trans}(b-d) for different directions of the light
propagation, and different $r/a$ ratios between the column radius $r$
and the lattice period $a$. Even if the refractive index contrast in
the photoresist-air PhC structures is too low for the formation of
photonic bandgaps, they can serve as templates for in-filling by
other materials with higher refractive index.

\begin{figure}[t]
\begin{center}
\includegraphics[width=12.50cm]{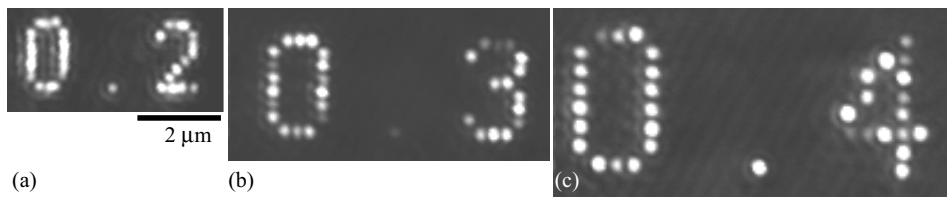}
\caption{\small{Readout of a ``bit'' pattern recorded in the volume
of silica at 10~$\mu$m depth by 150~fs pulses of 800~nm wavelength.
Readout wavelength was 488~nm. The distance between adjacent bits was
0.2, 0.3, and 0.4~$\mu$m for (a), (b), (c), respectively.}}
\label{f-gido}
\end{center}
\end{figure}
\subsection[Direct laser writing]{Direct laser writing}

Figure~\ref{f-gido} demonstrates optical readout of the 3D pattern
recorded by femtosecond laser pulses of the fundamental wavelength in
fused silica. Every bit was recorded in a single shot. The
diffraction-limited spot radius (radius of I$^{\textrm{st}}$ Airy
disk) during the pattern recording was about 349~nm (Rayleigh
criterion of resolution), whereas in the readout it was about
$0.61\lambda_{r}/NA = 213$~nm for $\lambda_{r}=488$~nm and $NA=1.4$.
Since the recording involved nonlinear absorption processes, the
recorded ``bits'' were considerably smaller than the focal spot
(diameter) of 697~nm of the recording pulse. Due to this
circumstance, 3D optical recording is possible with bit dimensions
$<0.2~\mu$m, and astonishingly high information density of
$BitVolume^{-1}=1/(0.2^{3}~\mu$m$^{3})=125~$Tbits/cm$^{3}$ can be
expected (1~Tbit$=10^{12}$~bits).

\section[Conclusions]{Conclusions}

We have demonstrated holographic recording of 3D structures by
interference patterns of multiple ultrashort laser pulses and direct
laser writing by tightly focused laser pulses. Typical feature size
achieved by these techniques can be easily decreased below the
diffractive limit ($\approx \lambda$) owing to nonlinear mechanisms
involved in the photmodification process. We have demonstrated direct
laser writing in silica with information density of about
100~Tbits/cm$^{3}$.

\begin{acknowledgments}
This work was in part supported by the Satellite Venture Business
Laboratory of the University of Tokushima.
\end{acknowledgments}


\end{document}